\documentclass[pra,aps,noeprint,nourl,nolongbibliography,superscriptaddress]{revtex4-2}

\usepackage{graphicx}
\usepackage{dcolumn}
\usepackage{bm}
\usepackage{float}

\begin{document}

\preprint{APS/123-QED}

\title{Unveiling the growth mode diagram of GaSe on sapphire}

\author{Michele Bissolo}
\email{michele.bissolo@tum.de}
\affiliation{%
 Walter Schottky Institute and TUM School of Natural Sciences, Technical University of Munich, 85748 Garching, Germany 
}%
\author{Marco Dembecki}%
\affiliation{%
 Walter Schottky Institute and TUM School of Natural Sciences, Technical University of Munich, 85748 Garching, Germany 
}%
\author{Jürgen Belz}
\affiliation{%
 mar.quest $\vert$ Marburg Center for Quantum Materials and Sustainable Technologies and Department of Physics, Philipps-Universität Marburg, 35032 Marburg, Germany 
}%
\author{Jan Schabesberger}
\affiliation{%
 Walter Schottky Institute and TUM School of Natural Sciences, Technical University of Munich, 85748 Garching, Germany 
}%
\author{Max Bergmann}
\affiliation{%
 mar.quest $\vert$ Marburg Center for Quantum Materials and Sustainable Technologies and Department of Physics, Philipps-Universität Marburg, 35032 Marburg, Germany 
}%
\author{Pavel Avdienko}
\affiliation{%
 Walter Schottky Institute and TUM School of Natural Sciences, Technical University of Munich, 85748 Garching, Germany 
}%
\author{Florian Rauscher}
\affiliation{%
 Walter Schottky Institute and TUM School of Natural Sciences, Technical University of Munich, 85748 Garching, Germany 
}%
\author{Abhilash S. Ulhe}
\affiliation{%
 Walter Schottky Institute and TUM School of Natural Sciences, Technical University of Munich, 85748 Garching, Germany 
}%
\author{Hubert Riedl}
\affiliation{%
 Walter Schottky Institute and TUM School of Natural Sciences, Technical University of Munich, 85748 Garching, Germany 
}%
\author{Kerstin Volz}
\affiliation{%
 mar.quest $\vert$ Marburg Center for Quantum Materials and Sustainable Technologies and Department of Physics, Philipps-Universität Marburg, 35032 Marburg, Germany 
}%
\author{Jonathan J. Finley}
\affiliation{%
 Walter Schottky Institute and TUM School of Natural Sciences, Technical University of Munich, 85748 Garching, Germany 
}%
\author{Eugenio Zallo}
\email{eugenio.zallo@tum.de}
\thanks{These authors jointly supervised this work}
\affiliation{%
 Walter Schottky Institute and TUM School of Natural Sciences, Technical University of Munich, 85748 Garching, Germany 
}%
\author{Gregor Koblm\"uller}
\altaffiliation[Current address: ]{Institute of Solid State Physics, Technical University Berlin, 10623 Berlin, Germany}
\thanks{These authors jointly supervised this work}
\affiliation{%
 Walter Schottky Institute and TUM School of Natural Sciences, Technical University of Munich, 85748 Garching, Germany 
}%


\begin{abstract}

The growth of two-dimensional epitaxial materials on industrially relevant substrates is critical for enabling their scalable synthesis and integration into next-generation technologies. Here we present a comprehensive study of the molecular beam epitaxial growth of gallium selenide on 2-inch c-plane sapphire substrates. Using \textit{in-situ} reflection high-energy electron diffraction (RHEED), \textit{in-situ} Raman spectroscopy, optical and scanning electron microscopies, we construct a diagram of the  gallium selenide growth modes as a function of substrate temperature (530–650~$^\circ$C) and Se/Ga flux ratio (5–110). The  growth mode diagram reveals distinct regimes, including the growth of layered post-transition metal monochalcogenide  GaSe with an unstrained in-plane lattice constant of 3.71$\pm$0.01~$\dot{\textrm{A}}$ and a partial epitaxial alignment on sapphire. 
This work demonstrates a RHEED-based pathway for synthesizing  gallium selenide of specific phase and morphology, and the construction of a phase diagram for high vapor pressure III-VI compounds that can be applied to a wide range of other metal chalcogenide materials.


\end{abstract}

\maketitle


\section{\label{sec:introduction}Introduction}

Gallium selenides, GaSe and Ga$_2$Se$_3$, are important III-VI semiconductors that have shown promising properties for a variety of electronic and optoelectronic applications.
While Ga$_2$Se$_3$ has been mainly explored for applications in photo-physics and non-linear optics~\cite{Guler_Ga2Se32019,Guo_Ga2Se32019}, GaSe has gained significant attention in recent years. As a two-dimensional (2D) van der Waals (vdW) material it belongs to the post transition metal monochalcogenide family (PTMCs, III-Se/S/Te) that has demonstrated promising (photo)electrochemical properties, excellent photoresponse, and viability in field-effect transistors with high on/off current ratios~\cite{cai_synthesis_2019,yu_review_2024, wang_high-performance_2015, shiffa_waferscale_2023,Sorifi_photodet2020,Demissie2024}. In particular, GaSe stands out as a very promising material for optoelectronics, owing to its tunable bandgap and the flat bands observed at the direct-to-indirect bandgap transition~\cite{zolyomi_band_2013,cao_tunable_2015}. These properties, along with a strong second harmonic generation (SHG) response~\cite{zhou_strong_2015}, make GaSe a compelling candidate for future technologies.
Ga$_{2}$Se$_{3}$ is a three-dimensional material with a defect zincblende structure in the cubic $\alpha$-phase, and it crystallizes in various vacancy-ordered polytypes, such as the monoclinic $\beta$- and orthorhombic $\gamma$-phases~\cite{Ho_Ga2Se32020}. GaSe, by contrast, is a layered material composed of monolayers built from Se-Ga-Ga-Se tetralayers arranged in a hexagonal lattice, with strong in-plane covalent bonds and weak van der Waals interactions out-of-plane. It crystallizes in various polytypes, which are defined by the stacking sequence of the tetralayers (e.g., $\beta$-, $\epsilon$-, $\delta$-, and $\gamma$-GaSe)~\cite{Grzonka_GaSepoly2021}.
Despite its intriguing features, realizing GaSe based materials and devices presents several challenges. Mechanical exfoliation, a commonly used method to obtain crystal layers, is not scalable due to its limited reproducibility and low yield of large-area materials. Furthermore, GaSe is highly susceptible to oxidation when exposed to air~\cite{Smiri2024}, leading to degradation of its (opto)electronic properties and limiting its practical applications.
To overcome these challenges, dedicated deposition methods are required, preferably performed under ultra-high vacuum (UHV) conditions, that allow \textit{in-situ} passivation to mitigate oxidation and improve material stability.
Molecular beam epitaxy (MBE) provides such desired UHV environment for the synthesis of thin PTMCs films~\cite{chen_large-grain_2018,zallo_two-dimensional_2023,song_wafer-scale_2023, shiffa_waferscale_2023}. The precise identification of the growth parameters yielding particular gallium selenide stoichiometries (1:1 vs 2:3) and surface morphologies (2D layered vs 3D) is especially vital for developing technologically relevant phases and scalable processes for high-quality gallium selenide-based devices. While much effort has been directed towards growing GaSe on Al$_2$O$_3$ (sapphire)~\cite{yu_epitaxial_2023, shiffa_waferscale_2023}, a cost-effective and industry-standard substrate with exceptional chemical and thermal stability, excellent electrical insulation, low dielectric loss, and wide optical transparency, the understanding of the different growth regimes is still lacking. First attempts in exploring different growth regimes of epitaxial GaSe by MBE were made~\cite{lee_molecular_2017}, yet the development of a comprehensive growth mode phase diagram that demarcates the specific growth regimes boundaries is still missing. \\
In this work, we present a wafer-scale investigation of the epitaxial growth parameters for gallium selenide on sapphire substrates by using solid-source MBE. A key aspect of our approach is the combination of dedicated \textit{in-situ} UHV monitoring techniques, in particular, reflection high-energy electron diffraction (RHEED) during deposition and \textit{in-situ} micro-Raman spectroscopy wafer mapping in an adjoining analytical UHV chamber.
This ensures that the gallium selenide surface remains free from oxidation and adsorbates throughout the analysis. It also allows us to characterize the as-grown material and systematically investigate how its cystalline phase and quality depend on growth parameters across complete 2-inch wafers.
GaSe was grown on multiple 2‑inch wafers, each presenting intrinsic spatial gradients in substrate temperature ($T_{sub}$) and VI/III flux ratio ($\Phi_{\text{Se/Ga}}$) due to the chamber and heater geometry, enabling exploration of a broad range of kinetic conditions within each individual sample.
By correlating these findings with the resulting morphology obtained by using scanning electron microscopy (SEM), optical imaging and atomic force microscopy (AFM), we identify several distinct growth regimes. These include the optimal window for achieving the desired 2D planar GaSe phase, as well as regimes for nanowire growth, the transition to the 2:3 stoichiometry, and high-surface-area gallium selenide 3D growth. This work sheds light on the fundamental aspects of gallium selenide growth far from equilibrium conditions and offers insights for improving its scalability and performance, bringing PTMC GaSe closer to widespread use in future optoelectronic devices.

\section{\label{sec:results}Results and Discussion}
\subsection{\label{sec:morph}Effect of substrate temperature and VI/III flux ratio}

Gallium selenide growth experiments were conducted by varying $\Phi_{\text{Se/Ga}}$ from 5 to 110 and $T_{sub}$ from 530 to 650~$^\circ$C, with each growth lasting one hour. The chamber’s geometry and sample manipulator design introduce significant gradients in the Ga flux and substrate temperature across the 2-inch wafers. These variations enable the exploration of a broad phase space within a single growth run, and by accurately determining these distributions (see the Methods) we can directly map the boundaries between different growth regimes. By assuming that varying the VI/III ratio via changes in either Ga or Se fluxes produces approximately equivalent effects, we can map the wafer-scale growth onto the growth mode phase diagram of gallium selenide. In the following, we describe the main information obtained by sampling the parameter space.
The data is presented logically from lowest to intermediate to highest investigated $\Phi_{\text{Se/Ga}}$.

\begin{figure*}[!htbp]
\includegraphics[width=\textwidth]{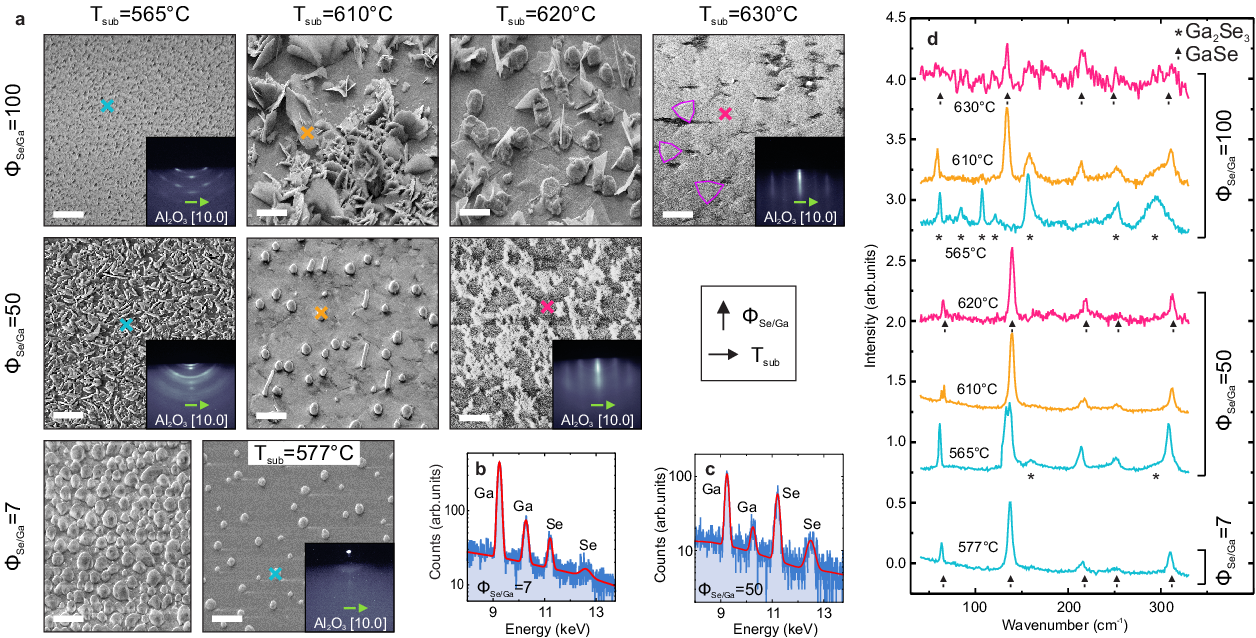}
\caption{\label{fig:123} 
Surface morphology and crystal phase as a function of $\Phi_{\text{Se/Ga}}$ and $T_{sub}$.
(a) SEM images showing the surface morphology of gallium selenide films grown at varying $\Phi_{\text{Se/Ga}}$ (rows) and $T_{sub}$ (columns), with selected RHEED patterns (inset) acquired perpendicular to the [10.0] direction. Scale bar: 2~µm. Crosses mark the positions where Raman measurements were performed.
(b-c) EDX spectra confirming the elemental composition of clusters on samples grown at $\Phi_{\text{Se/Ga}}$ of 7 (b) and 50 (c). The red lines are Gaussian fits to the experimental data. 
(d) Raman spectra from the marked regions in (a), identifying the predominant material phases (GaSe, Ga$_2$Se$_3$) under different growth conditions. Spectra are color-coded by growth temperature (cyan to magenta for low to high temperatures, as indicated in (a)). The Raman peaks were identified according to literature values \cite{Allakhverdiev_Raman2006, Molas_Raman2021, Yamada_Ga2Se31992}.}
\end{figure*}

\begin{figure*}[ht!]
\includegraphics[width=\textwidth]{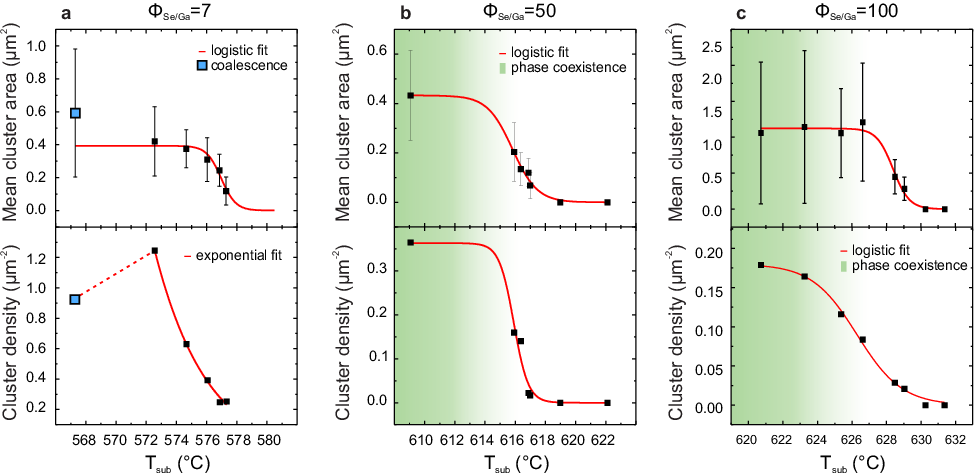}
\caption{\label{fig:4}
Area and density of clusters.
Cluster footprint area (top row) and density (bottom row) extracted from SEM images are shown as a function of $T_{sub}$ for three $\Phi_{\text{Se/Ga}}$ (from left to right: 7, 50, and 100). In (a), clusters primarily consist of Ga droplets and coalesce into larger sizes at high densities (light blue marker). In (b-c), clusters are highly selenidized and coexist with the 3D-nanoflakes morphology at lower $T_{sub}$ (light green region). To determine the onset temperature of cluster formation, the data is fitted (continuous red lines) with logistic functions (exponential in the bottom graph in (a)). Note that the actual onset values are extracted at $\Phi_{\text{Se/Ga}}$ of 8.3, 34.6, and 78.9 due to the additional Ga flux gradient on the wafer.
}
\end{figure*}

In Figure~\ref{fig:123}, SEM images depict the surface morphology of gallium selenide grown at $T_{sub}$ between 565 and 630~$^\circ$C with varying $\Phi_{\text{Se/Ga}}$ (7, 50 and 100). The surface shows a droplet-like morphology at the lowest $\Phi_{\text{Se/Ga}}$ of 7 (bottom row in Figure~\ref{fig:123}a), indicating Ga-rich growth conditions. At 565~$^\circ$C the droplets coalesce and almost fully cover the surface, while at higher $T_{sub}$ (577~$^\circ$C) their density and size are reduced due to the higher desorption rates of Ga (see details in the discussion of Figure~\ref{fig:4}a). In fact, the absence of a distinct RHEED pattern (see inset) suggests an amorphous or liquid-like surface structure~\cite{Koblmueller_InN2007}. The presence of Ga droplets in this growth regime is confirmed by the excess Ga content from the EDX data (see Figure~\ref{fig:123}b). However, the Raman spectrum (see bottom spectrum for $\Phi_{\text{Se/Ga}}$=7 in Figure~\ref{fig:123}d)) displays prominent characteristic peaks of the 1:1 GaSe stoichiometry, suggesting at least partial selenidization of the droplets. No signature of gallium selenide can be detected in the Raman spectra between the droplets. A further increase of $T_{sub}$ results in a reduction in droplet size, with, ultimately, no growth at higher $T_{sub}$ (see later in the text for the exact value extracted).


At 565~$^\circ$C, as $\Phi_{\text{Se/Ga}}$ increases to 50 (middle row in Figure~\ref{fig:123}a), the surface becomes textured, featuring irregular, nanoflake ensemble structures (from now on referred to as 3D-nanoflakes). Similar morphologies have been demonstrated to play an essential role in energy storage applications~\cite{xia_recent_2018}. The RHEED pattern (see inset) displays rings consistent with the random orientation of the crystallites. Raman spectroscopy ($\Phi_{\text{Se/Ga}}$=50 in Figure~\ref{fig:123}d) shows the characteristic peaks of GaSe as the dominant features, but peaks corresponding to the 2:3 Ga$_{2}$Se$_{3}$ stoichiometry are also visible. 
As $T_{sub}$ increases from 565~$^\circ$C to 600~$^\circ$C, the 3D-nanoflakes transition toward a mix of three distinct morphologies, with droplet-like features and 3D-nanoflakes now coexisting atop a smoother, triangular planar base layer (see left panel in Figure S1 in Supplementary Information). Further increasing $T_{sub}$ to 610~$^\circ$C (middle row, second image of Figure~\ref{fig:123}a) causes a suppression in 3D-nanoflakes growth and the 2D base layer remains intact. Interestingly, we observe a notable emergence of nanowires, as already reported for the growth of GaSe on GaAs~\cite{Sorokin_GaSeGaAs2020}. These extend from some droplets, suggesting a vapor-liquid-solid (VLS) growth process similar to that of III-V nanowires~\cite{Rudolph_NW2014, Ambrosini_GaAsNW2011, Li_GaAsSbNW2023}. In some cases, droplets atop the nanowires provided additional evidence of this growth mechanism (see Figure S1).
EDX data (Figure~\ref{fig:123}c) of the droplets at this $T_{\text{sub}}$ suggests a higher Se content compared to droplets formed at lower $T_{\text{sub}}$ and $\Phi_{\text{Se/Ga}}$, with the Ga:Se ratio approaching 1. However, residual effects from the planar base layer and geometric effects due to the small size of the clusters might influence these results, preventing the precise extraction of the stoichiometry and the confirmation of complete selenidization.\\
Finally, the increase of $T_{sub}$ to 620~$^\circ$C suppresses the droplet formation, leaving only the triangular planar features indicative of 2D GaSe growth, as also confirmed by the streaky RHEED (see inset). Raman spectra of Figure~\ref{fig:123}d ($\Phi_{\text{Se/Ga}}$=50) reveal that all three morphologies in the full $T_{\text{sub}}$ range (565--650~$^\circ$C) exhibit a nearly pure GaSe phase.

Furthermore, we performed a similar analysis at the highest $\Phi_{\text{Se/Ga}}$ of 100. As shown in the top left corner of Figure~\ref{fig:123}a, the surface smoothens significantly at $T_{\text{sub}}$=565~$^\circ$C, suggesting more uniform and planar growth. Interestingly, the RHEED pattern in the inset exhibits spots arranged in concentric rings
~\cite{hafez_review_2022}, indicating some degree of island growth and polycrystallinity. As confirmed by Raman spectroscopy ($\Phi_{\text{Se/Ga}}$=100 in Figure~\ref{fig:123}d), the sample consists only of 3D Ga$_2$Se$_3$. At 610~$^\circ$C, the gallium selenide surface undergoes a transition from this smooth, planar Ga$_2$Se$_3$ morphology to 3D-nanoflakes, similar to the case observed at lower $\Phi_{\text{Se/Ga}}$ and $T_{sub}$. The \textit{in-situ} Raman spectrum of Figure~\ref{fig:123}d ($\Phi_{\text{Se/Ga}}$=100) taken on such a nanostructure confirms a compositional shift, displaying spectral features characteristic of GaSe with a decreased Ga$_2$Se$_3$ signature.
As $T_{sub}$ increases to 620~$^\circ$C (top row of Figure~\ref{fig:123}a), the 3D-nanoflakes diminish in density whereas prominent clusters become the main surface feature. These clusters exhibit a partially faceted shape with flat top surfaces. Some clusters feature a Ga droplet at their peak, indicating VLS growth (see Figure S1 in  Supplementary Information). Raman spectra collected from the clusters show a mixed composition of both GaSe and Ga$_2$Se$_3$ phases, while EDX data is comparable to the spectrum observed for the droplets in Figure~\ref{fig:123}b (see Figure S2 and S3 in the Supplementary Information).
At the highest $T_{sub}$ of 630~$^\circ$C the surface morphology returns to a planar form, consisting of closely packed 2D rounded triangular features approximately 1.5~$\mu$m in size (see highlighted shapes from top row of Figure~\ref{fig:123}a). The RHEED pattern (inset) exhibits pronounced streaks aligned with the sapphire's [10.0] direction, confirming high crystalline quality. Additionally, weaker features corresponding to the [11.0] orientation of GaSe suggest some degree of in-plane disorder and misorientation relative to the sapphire substrate. A detailed discussion of the grain orientation and its implications will follow in section~\ref{sec:2Dphase}.


Since the formation of well-defined 2D triangular features at 630~$^\circ$C suggests that GaSe growth remains stable at this temperature, we explored whether higher T$_{\mathrm{sub}}$ could further improve the crystalline quality. However, at temperatures exceeding those that support planar growth (650~$^\circ$C, see section \ref{sec:2Dphase}), GaSe begins to desorb. This temperature threshold depends on $\Phi_{\text{Se/Ga}}$, as higher $\Phi_{\text{Se/Ga}}$ likely supports increased growth rates. Under these conditions, GaSe rapidly degrades after the cell shutters are closed, leading to its complete evaporation and leaving no residual material on the surface. Note that the SEM images represent regions near the transition between cluster formation, continuous film and desorbed regimes, and are intended to illustrate the morphological evolution across the growth mode phase diagram rather than optimized, fully planar films. At higher temperatures and flux ratios, GaSe tends to coalesce more effectively, and the area in parameter space exhibiting continuous coverage becomes larger. This trend is reflected in the measured surface morphology as a function of $T_{\mathrm{sub}}$ at a $\Phi_{\text{Se/Ga}}$ of 87 (see Section \ref{sec:2Dphase} for a detailed analysis of the optimal growth conditions for 2D morphology).\\
In order to describe the observed transition between cluster and 2D-like growth, we propose that Ga droplets play a vital role in this growth dynamic behavior. For example, as $\Phi_{\text{Se/Ga}}$ and $T_{sub}$ increase, the droplets become progressively more selenidized, as testified by a shift from a Ga-rich to Se-rich composition. The formation of clusters is consistent with prior studies of gallium selenide growth on sapphire, which attributed it to the high surface mobility of Ga, resulting in the development of Ga droplets and Ga$_2$Se$_3$ clusters~\cite{yu_epitaxial_2023}. The low Se concentration, caused by its low sticking coefficient~\cite{liu_sticking_2021}, the low reactivity due to the low Se cracking efficiency and sparse Se molecules, and the possible formation of volatile subselenides~\cite{Vogt_subselenides2025}, allow Ga to coalesce into droplets before reacting with Se. Here, we observe in the Raman and EDX data (see Figure~\ref{fig:123} and Figures S2 and S3 in Supplementary Information) that the cluster composition strongly depends on $\Phi_{\text{Se/Ga}}$, which is consistent with the progressive selenidization of Ga droplets.\\

To investigate the boundary of the droplet regime, we analyzed the density and footprint area of the clusters extracted from SEM images measured at various surface positions corresponding to different $T_{sub}$ (see Figure~\ref{fig:4}). For all examined $\Phi_{\text{Se/Ga}}$ (7, 50, 100), the cluster area exhibits a consistent trend: it remains constant at lower $T_{sub}$ and decreases exponentially at higher $T_{sub}$, as shown in the top graphs in Figures~\ref{fig:4}a-c (the error bars indicate the standard deviation of the cluster area). Similar patterns are observed for cluster density (bottom graphs in Figure~\ref{fig:4}), except at low $\Phi_{\text{Se/Ga}}$, where the density drops at lower $T_{sub}$.
This behavior is attributed to the coalescence of the droplets at high densities, where Ga droplets merge to form larger clusters, as evidenced by the increasing mean cluster area, leading to greater area variance as some droplets merge while others remain unaffected. At higher $\Phi_{\text{Se/Ga}}$ (Figures~\ref{fig:4}b-c), no coalescence is observed due to the earlier onset of parasitic 3D-nanoflakes growth. We attribute the saturation of cluster area and density, as well as the increase in area variance (see error bars in the top graphs of Figures~\ref{fig:4}b-c), to a shift in growth kinetics that promotes the formation of 3D-nanoflakes over clustering. The increased prevalence of parasitic 3D growth provides additional bonding sites for both Ga and Se, thereby suppressing Ga coalescence. 
This effect is observed at low $T_{sub}$, as the increased sticking coefficient of Se and lower Ga mobility limit the coalescence of Ga adatoms. 
Accordingly, 2D growth is facilitated over clustering at high $T_{sub}$, due to the increased Ga volatility. The Ga does not coalesce and either desorbs or reacts with the incoming Se. Interestingly, we observe a shift of the curves toward higher $T_{sub}$ as we increase $\Phi_{\text{Se/Ga}}$ (see Figure~\ref{fig:4}b,c). In order to maintain a sufficient Se surface coverage that reacts with both Ga and subselenides on the surface, more Se must be supplied so that the desorption is compensated. This increase in Se drives the VLS growth, stabilizing the cluster formation at higher temperatures.
To model this behavior and identify the cross-over temperature of these boundaries in the growth mode phase diagram, the curves were approximated using logistic functions. We note that these fits carry no physical meaning but help to effectively approximate the observed trends and capture both the exponential onset and eventual saturation of the cluster area and density. We define the onset temperature of cluster formation, $T_{\text{onset}}$, as $T_0 - 2/k$, where $T_0$ and $k$ are, respectively, the midpoint and growth rate of the logistic curve. For $\Phi_{\text{Se/Ga}}$ of 7, 50, and 100, $T_{\text{onset}}$ values are 577.6~$^\circ$C, 616.9~$^\circ$C, and 628.9~$^\circ$C, respectively. These onset temperatures will be used to construct the boundary demarcating the region that supports cluster formation (see section~\ref{sec:phasediagram}).

The characterization of the phase and morphological boundaries as a function of growth parameters provides a foundational understanding of the growth-phase dynamics across varying kinetic conditions. These will be combined in the next section with additional RHEED and Raman data to construct a comprehensive phase diagram of the growth regimes by mapping the morphology and crystalline phase transitions of gallium selenide on sapphire.

\subsection{\label{sec:phasediagram} Growth mode} phase diagram

\begin{figure*}[ht!]
\includegraphics[width=\textwidth]{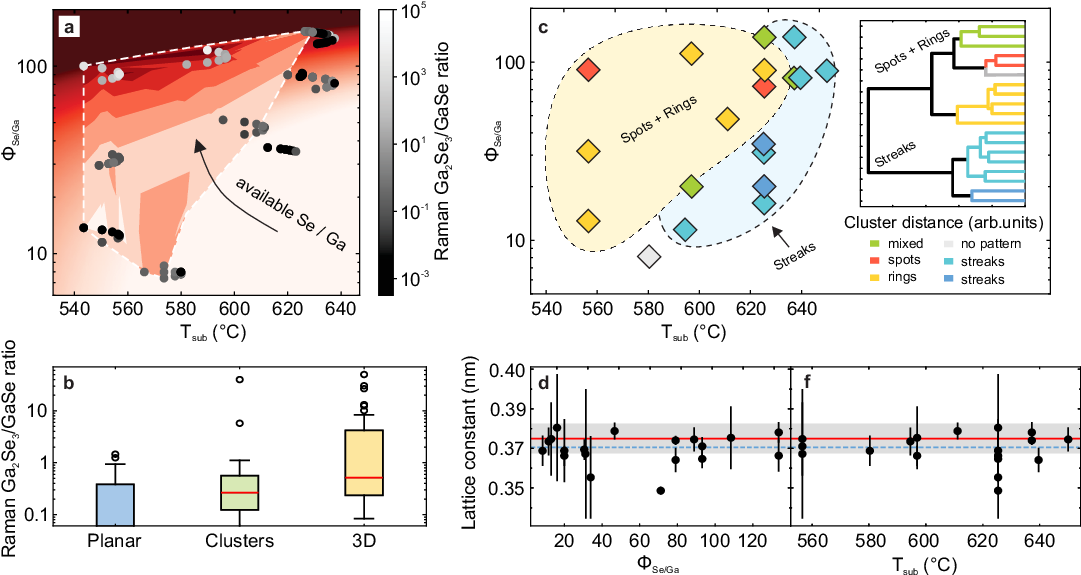}
\caption{\label{fig:5} 
\textit{In-situ} Raman and RHEED of gallium selenide samples.
(a) Scatter plot of Raman measurement positions in parameter space, with the marker color indicating the Ga$_2$Se$_3$ to GaSe ratio. The filled contour plot bounded by the white-dashed line also shows the Ga$_2$Se$_3$ to GaSe ratio, calculated only for samples where gallium selenide grows in a 3D mode. The background heatmap provides a qualitative extrapolation of the contour plot (refer to Supplementary Information Note S1 for the employed equation). The colorbar is omitted for both the contour plot and background heatmap as these are intended to illustrate trends, with quantitative values indicated by the marker colors.
(b) Boxplot of Ga$_2$Se$_3$ to GaSe ratios derived from Raman data, categorized by three distinct morphologies. The red line indicates the sample's median, while the box represents the interquartile range (IQR). The whiskers extend up to 1.5 times the IQR, with outlier points shown as white dots.
(c) Scatter plot of the grown samples as a function of $T_{sub}$ and $\Phi_{\text{Se/Ga}}$, with markers color-coded by RHEED pattern type (e.g. streaky, spotty+rings). The classification of RHEED patterns was performed algorithmically using a Histogram of Oriented Gradients (HOG)-based method (see main text). The accompanying dendrogram illustrates the clustering of similar RHEED patterns based on shape features: streaky patterns (light blue, indicative of 2D planar growth) and spotty/ring patterns (light yellow, associated with mixed or 3D growth).
(d) Scatter plot of lattice constant as a function of $T_{sub}$ and $\Phi_{\text{Se/Ga}}$, extracted from RHEED after 3 minutes of growth for all runs. The red line represents the literature lattice constant of GaSe, with the gray box showing a $\pm$0.5\% deviation. The dashed blue line indicates the mean lattice constant across all samples.}

\end{figure*}

To further investigate the structural composition of the grown material across different $T_{sub}$ and $\Phi_{\text{Se/Ga}}$ conditions, we analyzed Raman spectra from 9 growth runs, each including linescans along the sample surface at multiple measurement points. Figure~\ref{fig:5}a presents a scatter plot showing the positions of the samples in phase space, where each marker's color represents the ratio between the 2:3 and 1:1 stoichiometries by a Lorentzian fit of the GaSe $A'_1$ mode at 139.5$\pm$1.6~cm$^{-1}$~\cite{Allakhverdiev_Raman2006, Molas_Raman2021} and the Ga$_2$Se$_3$ $A_1$ mode at 162.7$\pm$2.6~cm$^{-1}$~\cite{Yamada_Ga2Se31992}. 
Overall, we observe that high $T_{sub}$ and low $\Phi_{\text{Se/Ga}}$ generally promotes a higher proportion of GaSe, while increasing $\Phi_{\text{Se/Ga}}$ at low (high) $T_{sub}$ favors the growth of Ga$_2$Se$_3$(GaSe). This trend is consistent with the previously observed temperature- and flux-dependent shift of the growth mode boundaries (see section~\ref{sec:morph}) and is ascribed to the varying availability of both Se and Ga on the surface as a function of $\Phi_{\text{Se/Ga}}$ and $T_{sub}$. 
By comparing the surface morphologies of Figure~\ref{fig:123}a with the relative amounts of GaSe and Ga$_2$Se$_3$ of Figure~\ref{fig:5}a, it is evident that the 2:3 stoichiometry is preferred only under conditions of high $\Phi_{\text{Se/Ga}}$ and low $T_{sub}$. This trend is further highlighted in the heatmap of Figure~\ref{fig:5}a, which is based on spectra collected exclusively from the 3D growth morphologies (3D-nanoflakes and smooth Ga$_2$Se$_3$). The filled contour map, enclosed by the dashed white line, represents the logarithm of the measured ratio between the 2:3 and 1:1 $A'_1$ Raman peaks (extrapolated outside the white dashed boundaries as a guide to the eye using the equation in Note S1 in Supplementary Information).  


The \textit{in-situ} spectroscopy allows us to probe the Raman spectra of different morphologies and gather statistical data. By quantifying the relative amounts of GaSe and Ga$_2$Se$_3$ for each of the three primary gallium selenide morphologies described in section~\ref{sec:morph}, we observe that the 2D planar growth consistently shows the lowest ratio of parasitic Ga$_2$Se$_3$. In contrast, the clusters and 3D-nanoflakes show a higher average Ga$_2$Se$_3$ content (see Figure~\ref{fig:5}b), confirming the results obtained from single spectra discussed in section~\ref{sec:morph} (note that the box labeled "3D" combines data from both the 3D-nanoflakes regions and the smooth GaSe observed at high fluxes and low temperatures). The clusters tend to form with a mixed composition with a preferred 1:1 stoichiometry (median ratio of 0.27). The distribution of the ratio for the 3D morphologies reflects the gradient in the filled contour plot of Figure~\ref{fig:5}a, showing significant variations across the phase space but a preference for the 1:1 stoichiometry (median ratio of 0.51) up to the crystalline phase transition, where at sufficiently high fluxes nearly pure Ga$_2$Se$_3$ forms.  
In regions where the GaSe $A'_1$ Raman peak is present, its full-width at half maximum (FWHM) as a function of growth parameters was extracted as an indicator of crystalline quality. The measured values are consistent with those obtained from exfoliated GaSe and reported in Figure S4 in the Supplementary Information.


These transitions are also evident in the RHEED patterns collected at the end of the growth, as visualized in Figure~\ref{fig:5}c by using the same axes as Figure~\ref{fig:5}a. In fact, the RHEED information serves as an effective real-time indicator of the surface morphology, providing immediate insight into the specific growth regime within the gallium selenide growth mode diagram. In order to avoid subjective biases in the manual classification of the data, we employ an algorithm based on the Histogram of Oriented Gradients (HOG)~\cite{Dalal_HOG2005}, which extracts information on the gradient orientation and magnitude of sub-regions of an image (edges). The HOGs for the different samples are then compared by calculating the distances and clustered according to Ward’s linkage method~\cite{Ward_linkage1963} (see dendrogram in the inset of Figure~\ref{fig:5}c).
At both low $T_{sub}$ and low $\Phi_{\text{Se/Ga}}$, the RHEED pattern appears dark and featureless (gray marker), indicating the accumulation of Ga droplets on the surface. This observation is consistent with the SEM and EDX results (see Figures~\ref{fig:123}a and b). Streaks dominate the highest $T_{sub}$ regions (blue data in Figure~\ref{fig:5}c), as expected from the planar morphology observed in the SEM images. At lower temperatures and higher $\Phi_{\text{Se/Ga}}$ (orange data in Figure~\ref{fig:5}c), the pattern shows clear rings, as confirmed by the randomly oriented 3D growth in the out-of-plane direction, a feature characteristic of disordered or polycrystalline surfaces. Samples grown near the growth mode boundary between 3D and 2D growth display a combination of both patterns (green markers) due to the large area probed by RHEED on the wafer (see Figure S5 in Supplementary Information).  

\begin{figure}[ht!]
\includegraphics[width=0.45\textwidth]{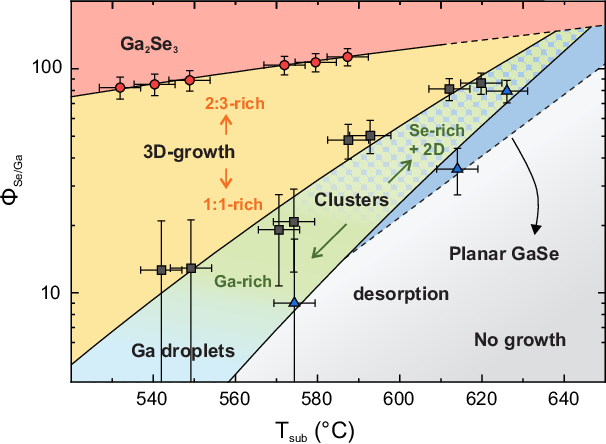}
\caption{\label{fig:6} 
Growth mode phase diagram of gallium selenide growth as a function of $T_{sub}$ and $\Phi_{\text{Se/Ga}}$. The predominant growth modes and crystalline phases are represented by color-coded growth regimes: yellow (3D-nanoflakes gallium selenide, predominantly 1:1 stoichiometry), red (Ga$_2$Se$_3$, 2:3 stoichiometry), light blue (Ga droplets), green with blue checks (clusters with GaSe underlayer), green (clusters with no underlayer), blue (2D GaSe growth), and white/gray (desorption or no growth).
Data points with error bars represent the experimentally measured boundaries between growth modes. Solid lines are Arrhenius fits corresponding to the experimentally derived boundaries. The boundaries defined by the red and dark gray markers are based on correlations between SEM and optical images. The transition extracted from the blue triangles is derived from SEM-based cluster density and footprint area measurements (see Figure~\ref{fig:4}).}

\end{figure}

Further analysis of the RHEED data at the beginning of deposition reveals that each growth run goes through a darker transition stage characteristic of amorphous growth (see Figure S6 in Supplementary Information). This is in contrast to the vdW-epitaxy, where the diffraction pattern smoothly transitions from substrate to grown material~\cite{Bissolo2025}. However, after a few minutes, the streaks show the formation of planar crystalline layers. It is important to note that this initial GaSe growth is followed by the onset of 3D growth at a later stage in some areas of the parameter space (see Figures~\ref{fig:123}a, \ref{fig:5}, and later \ref{fig:6}). Figure~\ref{fig:5}d shows the in-plane lattice constant for all samples obtained from the RHEED data at the beginning of the growth (3 min). No clear dependence on $\Phi_{\text{Se/Ga}}$ or $T_{sub}$ is observed, and the measured mean value of 3.71$\pm$0.01~$\dot{\textrm{A}}$ deviates by only 1\% from the literature value of 3.75~$\dot{\textrm{A}}$~\cite{Kuhn_crystalstructure1975, Li_GaSe_on_Gr2023}. This indicates that the strain accumulated at the interface is minimal and independent of the location in the examined portion of parameter space. Thus, in the early stages, the growth of gallium selenide on sapphire appears to be not strongly influenced by the diffusion dynamics and density of atomic species, which are both highly dependent on $\Phi_{\text{Se/Ga}}$ and $T_{sub}$. This also explains the large region of the parameter space that supports 2D growth (both with and without clusters). This behavior could arise from the formation of an interface layer that acts as a disordered buffer across the whole phase diagram, which supports planar GaSe growth in the initial stages of deposition. This view is also supported by the lack of material contrast between flakes and substrate in the indentation and adhesion maps of samples grown at high temperature, where the majority of the GaSe film desorbs at the end of the growth (see Figure S7 in Supplementary Information). We note that an amorphous interfacial layer has been reported previously for the growth of GaSe on sapphire~\cite{shiffa_waferscale_2023}, which motivated the more detailed investigation of the interface presented in Section~\ref{sec:2Dphase}.\\ 

Finally, by compiling all the information obtained from the $T_{sub}$- and $\Phi_{\text{Se/Ga}}$-dependent SEM, Raman, RHEED, and EDX measurements, as well as optical images of the full 2-inch wafer (see Figures S8 and S9 in Supplementary Information), we can construct a growth mode phase diagram, which is shown in Figure~\ref{fig:6}.
A large portion of the parameter space (yellow) is defined by non-oriented growth. Here, the grown gallium selenide appears as 3D-nanoflakes, mainly consisting of GaSe, with significant in-plane and out-of-plane misorientation (see also the SEM images in Figure~\ref{fig:123}). By visual inspection, the wafers appear dark red, typical of bulk gallium selenides, and hazy due to the highly texturized 3D surface. Increasing $\Phi_{\text{Se/Ga}}$ leads to a higher Ga$_2$Se$_3$ content within the film until the growth transitions to a smoother film primarily consisting of Ga$_2$Se$_3$ and characterized by a reduced out-of-plane misorientation (see the red area in Figure~\ref{fig:6} and RHEED transitions from rings to spots in Figure~\ref{fig:5}). The boundary between the 3D-nanoflakes growth and Ga$_2$Se$_3$ can be extracted from the optical images due to increased surface reflection and the change to a lighter shade of red.
In particular, the cross-over $T_{sub}$ and $\Phi_{\text{Se/Ga}}$ for the growth mode transitions can be obtained from the known distribution of growth parameters on the sample (see red markers in Figure~\ref{fig:6}).

The phase diagram becomes more complex when we move from the 3D region to higher temperatures. As $T_{sub}$ is raised, the 3D-nanoflakes transition into clusters, with a 2D layer observed between them (light green region; the blue checks indicate the areas of parameter space that support the GaSe underlayer). The optical appearance of the surface changes drastically, losing the characteristic red color and shifting to a hazy milky white, which is attributed to the diffuse scattering of the micrometer-sized clusters (see section~\ref{sec:morph}). This boundary is extracted from the optical images and shown in Figure~\ref{fig:6} by the dark gray markers. Further increase in $T_{sub}$ reduces the size and density of these clusters until only planar 2D triangles remain (see the blue region and section~\ref{sec:expdet2}). The blue markers indicate the boundary for cluster formation derived from the fits in Figure~\ref{fig:4}. The solid lines represent the  growth mode boundaries and are exponential fits to the data points with Arrhenius-like behavior. The fits return effective activation energies of 2.75, 1.82, and 0.36 eV for the blue, dark gray, and red transitions, respectively. We tentatively assign the 2.75~eV barrier to the energetic difference between Ga coalescence, selenidation and adatom desorption, the 1.82~eV barrier to the competition between adatom diffusion and direct nucleation, and the 0.36~eV barrier to the activation energy for stoichiometric conversion between GaSe and Ga$_{2}$Se$_{3}$. These assignments are, however, based solely on the growth phase diagram we constructed.  
The physical meaning of these activation energies remains unclear and may involve multiple competing mechanisms that are difficult to isolate within the current framework. For example, recent studies on III–Se epitaxy have proposed a two-step growth mechanism mediated by volatile subselenides (GaSe, Ga$_2$Se), rather than direct Ga + Se reactions, with these subcompounds acting as key cation-like intermediates that either desorb or undergo further selenidation to Ga$_2$Se$_3$~\cite{Vogt_subselenides2025}. Within this framework, our observations can be understood by considering that increasing Se flux converts GaSe and Ga$_2$Se into Ga$_2$Se$_3$, naturally defining the Ga$_2$Se$_3$/GaSe boundary.\\
In the cluster growth region (light green and light blue regions of the diagram in Figure~\ref{fig:6}), which appears to be driven by a VLS-type mechanism, Ga droplets accumulate on the surface and become progressively selenidized depending on $\Phi_{\text{Se/Ga}}$. At low $\Phi_{\text{Se/Ga}}$, the Ga droplets retain their liquid form (see the high Ga content in the EDX spectra of Figure~\ref{fig:123}d) while higher Se content leads to columnar and nanowire growth catalyzed by Ga droplets (see Figure~\ref{fig:123}a). Similarly, the 2D layer observed between the clusters is also dependent on $\Phi_{\text{Se/Ga}}$, with no 2D phase observed at low $\Phi_{\text{Se/Ga}}$ (see blue check pattern in the diagram of Figure~\ref{fig:6} and surface morphology in Figure~\ref{fig:123}a). Overall, the entire phase diagram exhibits a slanted shape, as visualized already by the analysis of the surface morphology, Raman maps, and RHEED in Figures~\ref{fig:123}a, \ref{fig:5}a, and \ref{fig:5}c, respectively. This is attributed to the temperature-dependent sticking coefficient of Se, the desorption rate of subselenides, which remove Ga from the surface, and the fast decomposition of gallium selenide~\cite{lee_molecular_2017, Vogt_subselenides2025}. As discussed in section~\ref{sec:morph}, as $T_{sub}$ increases, fewer Se and Ga diffuse across the surface due to rapid desorption from the substrate. Ga and subselenides can, in turn, incorporate Se, and the cluster growth is thus inhibited, leaving the 2D layer to dominate the growth mode. Therefore, we tentatively attribute the formation of the 2D GaSe layer to the free Se and Ga adatoms and intermediate subcompounds nucleating on the surface with no Ga droplet seeding the growth of gallium selenide (see section~\ref{sec:2Dphase} for more details).
At even higher $T_{sub}$ (see the dashed line separating the blue 2D planar phase from the white region), the GaSe film becomes unstable, leading to either rapid desorption after closing the Se shutter at the end of the growth or a decreased growth rate to the point where no material is deposited. Dedicated future studies, such as line-of-sight mass spectrometry to detect desorbing Ga-Se species or kinetic Monte Carlo simulations, could be used to identify the exact chemistry and kinetics that define these boundaries.

\subsection{\label{sec:2Dphase}2D G\lowercase{a}S\lowercase{e}}

The 2D growth regime of monochalcogenide GaSe with 1:1 stoichiometry holds particular significance due to its desirable properties for optoelectronic applications~\cite{shiffa_waferscale_2023,cai_synthesis_2019}. Nucleating and growing the material with this phase requires precise control over the growth parameters, as demonstrated by the growth mode boundaries mapped in this study (see the blue region in the phase diagram in Figure~\ref{fig:6}). In the following, we demonstrate that GaSe is partially aligned with the substrate and forms smooth and planar surface morphology within the 2D regime. It is important to note that we use the term '2D growth regime' here to refer to the van der Waals (vdW) 2D form of GaSe, rather than the layer-by-layer growth typical of 3D-bonded materials. 

\begin{figure}[ht!]
\includegraphics[width=0.45\textwidth]{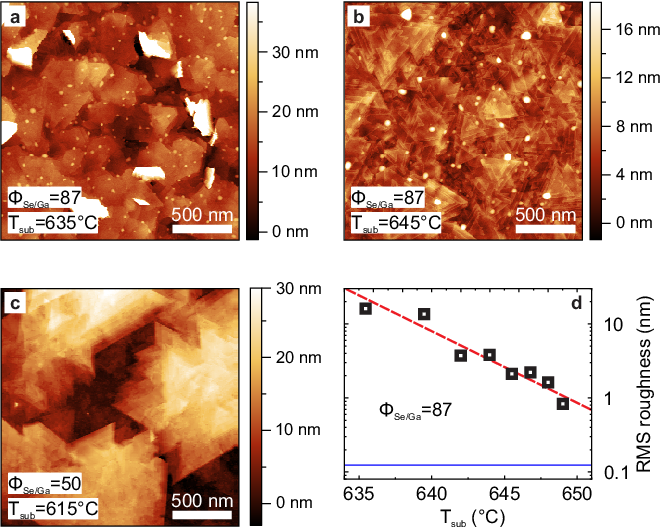}
\caption{\label{fig:7} 
AFM maps of the grown films. (a-b) Surface morphology of the GaSe films grown with $\Phi_{\text{Se/Ga}}$=87 at $T_{sub}$=635~$^\circ$C (a) and $T_{sub}$=645~$^\circ$C (b). (c) Surface morphology of GaSe grown with $\Phi_{\text{Se/Ga}}$=50 at 615~$^\circ$C. (d) Root-mean-square (RMS) roughness extracted from 4~$\mu m^2$ AFM micrographs as a function of $T_{sub}$.}

\end{figure}

To assess the effect of $T_{sub}$ and $\Phi_{\text{Se/Ga}}$ on the surface morphology, a series of AFM measurements was performed. Overall, we observe that lower temperatures lead to higher surface roughness (see Figure~\ref{fig:7}a,b). At a lower $T_{sub}$ of 615~$^\circ$C and $\Phi_{\text{Se/Ga}}$ of 50, the surface exhibits large triangular domains separated by deep trenches, indicating poor coalescence and resulting in a root-mean-square (RMS) roughness of 12.59~nm (Figure~\ref{fig:7}c). At 635~$^\circ$C with a higher $\Phi_{\text{Se/Ga}}$ of 87, the surface becomes more uniform with coalescence and smaller features, but still displays protruding structures that increase the roughness to 15.23~nm (Figure~\ref{fig:7}a). Increasing $T_{sub}$ further to 645~$^\circ$C, while keeping $\Phi_{\text{Se/Ga}}$ at 87, results in a flat and continuous morphology with well-defined triangular features and a significantly reduced roughness of 2.13~nm (Figure~\ref{fig:7}a), which is in line with the best values typically reported for this system~\cite{Tran_InGaSe-GaSe2024}. This trend is quantitatively captured in the RMS roughness of Figure~\ref{fig:7}d. The plot shows an exponential decrease in roughness with increasing $T_{sub}$, indicating that optimal coalescence and smoothness are achieved under high-temperature, Se-rich conditions (see Figure S10 for selected AFM micrographs).

\begin{figure}[ht!]
\includegraphics[width=\textwidth]{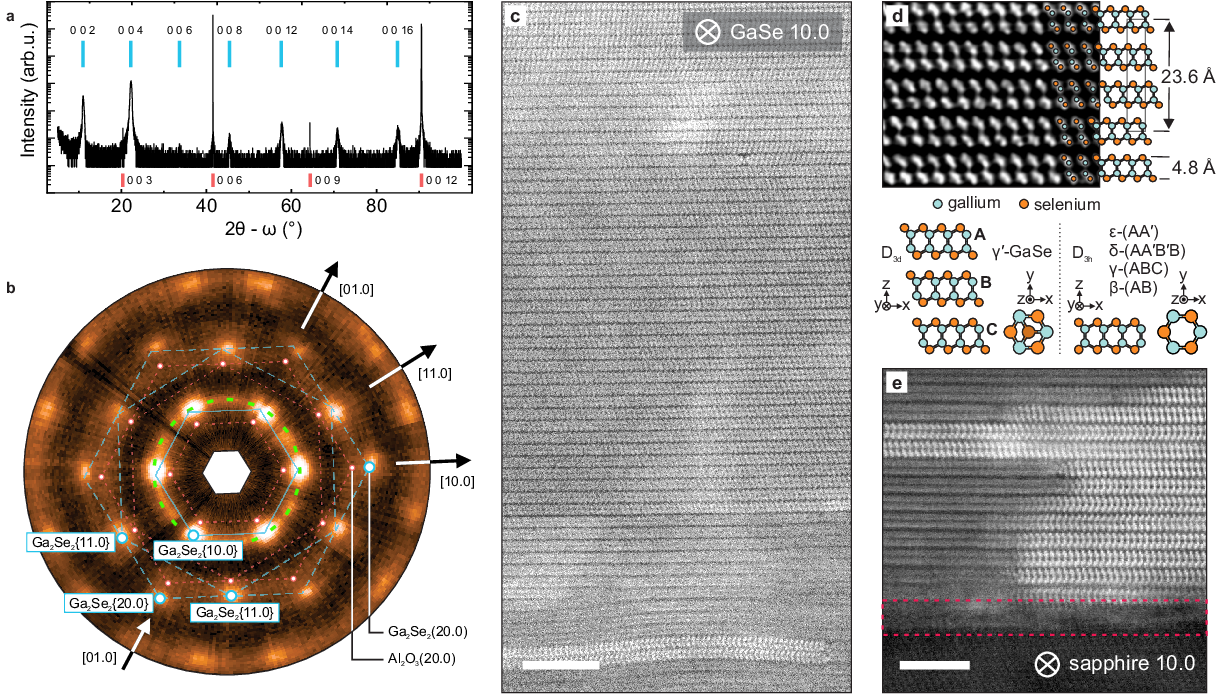}
\caption{\label{fig:8} 
Structural properties of the GaSe films. (a) HR-XRD 2$\theta$-$\omega$ scans of the GaSe film. The light blue (red) lines indicate the positions of the GaSe (sapphire) reflections. (b) In-plane azimuthal RHEED map of GaSe. The contributions from GaSe and Al$_2$O$_3$ (the second not discernible in the RHEED pattern itself) have been marked with light blue and red circles. The dashed green line indicates the azimuthal linescan from which the intensity distribution shown in Figure S13 is extracted. (c) Cross-sectional HR-STEM image of a 87 nm-thick GaSe layer recorded along the [10.0] direction. Scale bar: 5 nm. (d) Filtered zoom-in STEM image showing the stacking of single GaSe layers overlayed with the $\gamma'$-GaSe crystal structure (top). Schematic models of $\gamma'$-GaSe (centrosymmetric tetralayer), and $\beta$-, $\epsilon$-, $\delta$-, and $\gamma$-GaSe (mirror-symmetric tetralayer). (e) STEM image of the GaSe/sapphire interface. The red dashed box highlights the fuzzy region between GaSe and sapphire. Scale bar: 3 nm.}

\end{figure}

Within this regime, GaSe grows phase-pure, as evinced by the HR-XRD 2$\theta$-$\omega$ scans of Figure \ref{fig:8}a (see also STEM-EDX data in Figure S11), with an out-of-plane lattice constant of 15.92~\AA, in perfect agreement with previous reports~\cite{Grzonka_GaSepoly2021}. A rocking curve measurement on the 006 peak reveals a FWHM of 0.4° (see Figure S12 in the Supplementary Information), indicating an overall good crystal quality, with additional tails that have been observed before in GaSe grown on GaAs(111)B and have been associated with a poor interface with the substrate~\cite{Yu_agedsubstrate2024} (see the discussion of the GaSe/sapphire interface below). As shown in Figure~\ref{fig:123}a, in this growth regime the 2D GaSe phase shows pronounced RHEED streaks, as GaSe is partially aligned with the sapphire's [01.0] direction. To further investigate the in-plane registry and symmetry of the epitaxial layer, an \textit{in-situ} azimuthal RHEED scan was conducted by measuring the streak intensities while varying the in-plane azimuthal angle as a function of the in-plane momentum transfer ($k_{\parallel}$). Figure~\ref{fig:8}b provides the complete static RHEED pattern recorded parallel to the surface, with the plane defined by the [10.0] and [11.0] vectors of the Al$_2$O$_3$ substrate, which was taken as a reference before growth. Red circles mark a few substrate reflections since they are not directly discernible in the RHEED pattern due to the thickness of the grown film.\\
The pole figure reveals sharp diffraction peaks every 60$^\circ$ corresponding to GaSe~$\{10.0\}$, $\{11.0\}$ and $\{20.0\}$ planes (red circles), confirming the expected hexagonal symmetry of GaSe. By analyzing the radial distribution of the peak intensity, we notice that GaSe is predominantly oriented epitaxially with the sapphire substrate~\cite{Bissolo2025, mortelmans_peculiar_2019}. However, a diffuse angle-independent background signal is observable in the \{10.0\} family of reflections (the innermost features). Based on the fit of the radial distribution of the \{10.0\} streaks (along the green dashed line in Figure~\ref{fig:8}; see Figure S13 in Supplementary Information), approximately 51\% of the GaSe surface is aligned with the sapphire substrate, while the remaining 49\% displays random in-plane orientation. Notably, no additional sub-domains were detected.

To correlate the surface and structural information with the interface quality and the crystalline stacking, and understand the source of the incomplete epitaxial alignment, cross-sectional high-resolution scanning transmission electron microscopy (HR-STEM) of a sample grown with $T_{sub}$=642~$^\circ$C and $\Phi_{\text{Se/Ga}}$=87 was performed. The overview image in Figure~\ref{fig:8}c evidences large areas with well-defined and uniform stacking. The magnified and filtered view reported in Figure~\ref{fig:8}d further reveals the characteristic $\gamma'$ stacking typical of MBE-grown GaSe~\cite{Yu_GaSeonGaAs2024, Grzonka_GaSepoly2021,shiffa_waferscale_2023}, consistent with the structural model shown alongside and a tetralayer thickness of 4.8~\AA, in good agreement with previous studies (note that the STEM image was calibrated using the lattice spacings of sapphire)~\cite{shiffa_waferscale_2023}. However, parts of the film show variations in crystallinity (see Figures~\ref{fig:8}c and S14 in the Supplementary Information) and a significant amount of disorder in terms of polytype with the coexistence of centrosymmetric and non-centrosymmetric tetralayer structures (see Figure~\ref{fig:8}e). In addition, Figure~\ref{fig:8}e reveals a distinct amorphous interfacial layer at the GaSe/sapphire interface, consistent with both azimuthal RHEED pattern (see Figure~\ref{fig:8}b) and PF-AFM observations, which likely disrupts the formation of a uniform epitaxial registry. The observed mixture of alignment and disorder could result from competing nucleation mechanisms, such as local pinhole-mediated nucleation or remote epitaxy~\cite{Henksmeier_remote-aC2024}, which may enable alignment in some regions, while other regions undergo vdW epitaxy with no crystallographic relationship to the substrate. Figure S14 (see Supplementary Information) shows extended lattice defects nucleating from this region, as well as a large-area STEM image with several GaSe grains merging together, supporting the understanding that the film evolves by coalescence of individual crystalline domains during growth.
\\

Similar amorphous interlayers have been observed previously for GaSe grown on sapphire~\cite{shiffa_waferscale_2023}. In comparison, GaSe grown on Si(111), GaAs, or GaN has been reported to form crystalline interfacial layers~\cite{Grzonka_GaSepoly2021, Yu_GaSeonGaAs2024, lee_molecular_2017}. The presence of the amorphous interlayer on sapphire may reflect the large lattice mismatch ($\sim$20\%) or the reduced surface reactivity of the substrate. By decoupling the film from the substrate, this amorphous layer can accommodate the large lattice mismatch and effectively suppress strain, as reflected by the in-plane lattice constant extracted from RHEED (Fig.~\ref{fig:5}d), invariant across a wide range of $T_{\text{sub}}$ and $\Phi_{\text{Se/Ga}}$. Consequently, the strain accumulated at the interface is independent of the explored growth parameters, suggesting that the formation of the interfacial layer is governed less by adatom diffusion and flux ratios and more by substrate-mediated interactions. Once this interfacial layer is established, GaSe growth becomes only partially coupled to the sapphire substrate. The disruption of uniform epitaxial registry gives rise to a mixture of well-aligned and misoriented domains, as the film evolves through domain nucleation and coalescence. In this regime, surface kinetics take over with diffusion-driven processes such as Ga clustering and 3D-nanoflake nucleation dominating the growth.

\section{Conclusions}

In this paper, we provided an understanding of the molecular beam epitaxial gallium selenide growth on sapphire substrates. Using a combination of \textit{in-situ} RHEED, \textit{in-situ} Raman spectroscopy, optical imaging and SEM microscopy, we have constructed a detailed growth mode phase diagram that delineates the crystalline phase transitions and morphological changes within the $T_{sub}$ range of approximately 530–650~$^\circ$C and $\Phi_{\text{Se/Ga}}$ from 5 to 110. By extracting the boundaries between growth modes from optical imaging and SEM microscopy data, we revealed the regions of parameter space that support the growth of GaSe and Ga$_2$Se$_3$ (1:1 and 2:3 Ga:Se stoichiometry) and their various corresponding morphologies. Specifically, we highlight the conditions for planar and 3D growth regimes, with GaSe forming at higher $T_{sub}$ and Ga$_2$Se$_3$ being stabilized by increased $\Phi_{\text{Se/Ga}}$.
The dependence of RHEED patterns on these growth regimes allows real-time monitoring of 2D-to-3D transitions during growth. 
The dynamical control of growth conditions based on the RHEED feedback demonstrates its utility as a diagnostic tool for optimizing the phase, morphology, and structural properties of GaSe films on sapphire, enabling efficient navigation of the phase diagram without interrupting the growth process or relying on \textit{ex-situ} techniques.
Our results show that the growth of gallium selenide on sapphire is characterized by the formation of a thin amorphous interfacial layer that accommodates the large lattice mismatch and minimizes strain, as confirmed by RHEED, PF-AFM, and HR-STEM. This interlayer provides effective passivation that enables 2D growth across a wide range of conditions, yet at the same time weakens the epitaxial relationship with the substrate, resulting in only partial in-plane alignment of the GaSe domains (51\% ). Optimized growth already yields smooth films with an RMS roughness of 2.13~nm, but strategies to improve epitaxial alignment and reduce or remove the amorphous interlayer could further minimize the roughness toward values observed in purely epitaxial systems~\cite{Tran_InGaSe-GaSe2024}. Future efforts could explore employing crystalline buffer layers (e.g. AlN) to decrease the mismatch and suppress interlayer formation, as well as tailoring the sapphire substrate by direct surface engineering (termination control, reconstruction, passivation) or through step-edge–assisted nucleation on vicinal surfaces, to modify the nucleation environment and promote unidirectional epitaxy~\cite{mortelmans_peculiar_2019,Li2021-hw}. Substrate patterning to confine growth to defined regions may further favour single-domain nucleation, improving both the alignment and roughness~\cite{kim_non-epitaxial_2023}. Moreover, conducting a systematic time-series study in the 2D growth regime to assess the effect of growth duration and resulting film thickness on surface texture and domain alignment would be a logical continuation of this work.\\
In conclusion, this study offers valuable insights into the parameter space of molecular beam epitaxial gallium selenide growth, serving as a tool to identify the growth conditions necessary to produce the desired gallium selenide composition and morphology. The approach also establishes a framework for optimizing the synthesis of other 2D PTMCs, such as InSe, SnSe, and GeSe, and advancing their integration into future technologies..

\section{\label{sec:methods}Methods}
\subsection{\label{sec:expdet}Experimental details}
The thin films of gallium selenide were grown in a RIBER MBE C21 system under a background pressure of 1.0×10$^{-9}$~mbar. High-purity Ga (7N) and Se (7N) were evaporated from a Veeco double-filament Knudsen cell and a RIBER VCOR valved cracker cell, respectively. The beam-equivalent-pressure (BEP) of both Ga and Se was measured with a Bayard-Alpert gauge at the position of the substrate before each growth run. The 2-inch one-side polished c-plane (0001) sapphire substrates were metallized on the backside with Ti/Pt (20/70~nm) for radiative heat transfer from the substrate heater.
Before growth, the substrates underwent a cleaning process involving sequential sonication in acetone, methanol, and isopropanol (IPA), followed by a nitrogen drying step after a final IPA rinse. Once loaded into the UHV system, the substrates were annealed at 450~$^\circ$C (thermocouple temperature) in a designated chamber at 1.0×10$^{-10}$~mbar for 12 hours to remove any adsorbed contaminants.
After annealing, the substrates were transferred to the growth chamber, where a thin sacrificial GaSe layer was deposited, followed by annealing at 820~$^\circ$C for 30 minutes. The desorption was monitored with RHEED to ensure the complete evaporation of the deposited layer. This process has been shown to enhance the subsequent epitaxial growth of GaSe on sapphire~\cite{chegwidden_molecular_1998, yu_epitaxial_2023}. At the end of the growth, the sample was first cooled to $\sim$350~$^\circ$C under a continued Se flux, then to room temperature with all cell shutters closed. The Ga-flux was kept constant at 6.9×10$^{-8}$~mbar (BEP) for all the samples, while the Se-flux was varied from 3.6×10$^{-7}$ to 7.4×10$^{-6}$~mbar. All growths were performed without substrate rotation and with the Se cell reservoir and cracker temperatures at 285~$^\circ$C and 700~$^\circ$C.

RHEED was used to monitor \textit{in-situ} the surface crystal structure in the center region of the 2-inch wafer before, during, and after growth. The probed area spans approximately 25 mm (1 inch) along the electron beam path and 2 mm perpendicular to it (see Figure S5 in Supplementary Information). Note that we use the four-component Miller-Bravais notation for the hexagonal symmetries of the sapphire and GaSe surfaces, i.e. (hklm) = (h k -(h+k) m) = (hk.m). The structural and optical properties of the resulting samples were extensively characterized by Peak-Force Atomic Force (PF-AFM), Scanning Electron (SEM) microscopies, Energy dispersive X-ray (EDX), and Raman spectroscopies. The PF-AFM maps were measured using a Bruker Dimension Icon AFM in peak force tapping mode. The SEM images were taken in a Zeiss NVision 40 SEM at a viewing angle of 45$^\circ$ and an accelerating voltage of 5~kV, while the EDX data was collected at 30~kV in a Zeiss EVO MA10 equipped with a Bruker XFlash detector.
Micro-Raman measurements were conducted in backscattering geometry right after the growth in a custom-made UHV chamber attached to the MBE growth system at room temperature by means of a continuous wave laser with 532 nm excitation wavelength (1.2 $\mu$m diameter spot) and spectral resolution $<$1 cm$^{-1}$.
Full-wafer optical images were captured in a nitrogen glovebox using a Nikon Z30 camera equipped with a Sigma 105mm macro objective and a circular polarizing filter. To minimize reflections, all images were taken at a slight angle from the wafer's surface normal. 
The cross-sectional, electron-transparent lamella was prepared using a JEOL JIB 4601 Ga-FIB, thinned on TEM grids and finally polished using Xe ions at 1 kV. STEM experiments were performed on a double aberration-corrected JEOL JEM-2200FS operating at 200 kV. The local composition was probed by energy-dispersive X-ray spectroscopy (EDX) using an XFlash 5060 detector.

\subsection{\label{sec:expdet2}Sampling parameter} space with flux and temperature gradients

Due to the chamber geometry and the design of the sample manipulator, the growth on 2-inch wafers is characterized by pronounced gradients in both the Ga flux distribution and the substrate temperature $T_{sub}$. While these variations in growth parameters across the wafer can negatively impact the spatial homogeneity of the films, they simultaneously allow us to probe a large region of the phase diagram in a single run and reduce the number of growth experiments needed to fully characterize the grown material. Additionally, as the exact flux and temperature distributions can be determined, the boundaries between growth regimes can be directly resolved.
Hereby, the Ga flux distribution on the substrate was estimated by using the formula~\cite{Herman2004}
\begin{equation} \label{eq}
\Phi_p = \frac{\Phi}{\pi r_p^2} \cos \theta \cos(\theta + \varphi),
\end{equation}

where \( \Phi \) is the flux density emitted from the source, and \( r_p \) is the distance between the center of the crucible orifice and a point \( p \) on the substrate. The angles \( \theta \) and \( \varphi \) are the effusion angle and the angle of the effusion cell relative to the substrate normal at the center, respectively.

The temperature distribution across the 2-inch wafer was measured through an optical-grade fused silica viewport using a Zelux 1.6 MP Monochrome CMOS Camera, which is sensitive to wavelengths up to 1100 nm (Figure S15). To minimize interference from stray light, a 700 nm long pass filter was placed in front of the camera.
The camera signal was calibrated based on the known central temperature of the 2-inch wafer. This central temperature was measured using a pyrometer, calibrated by placing an indium droplet at the center of a wafer and tracking its desorption with a line-of-sight mass spectrometer as the temperature changed. The desorption data were fitted to an Arrhenius relationship using the known indium heat of vaporization of 2.4~eV~\cite{barin_thermochemical_1995} to extract the precise temperature at the center. With this information and by not rotating the sample during deposition, we were able to map the structural and optical properties of the grown gallium selenide film directly onto the $T_{sub}$ versus $\Phi_{\text{Se/Ga}}$ space, as illustrated in Figure S16 (Supplementary Information).

\begin{acknowledgments}
The authors would like to thank R. Günkel for the insightful feedback and careful proofreading of the manuscript. The work was partly funded by the German Research Foundation (DFG) under Germany's Excellence Strategy via the Clusters of Excellence e-conversion (EXC No. 2089/1-390776260) and the MCQST (EXC-2111-390814868) and via Grants FI 947/7-1, FI 947/8-1 and KO4005-9/1. J.B., Ma.B., and K.V. acknowledge support from the German Research Foundation (DFG) within the Collaborative Research Center SFB 1083 (project no. 223848855) and from the European Regional Development Fund (ERDF) and the Recovery Assistance for Cohesion and the Territories of Europe (REACT-EU).

\end{acknowledgments}

\section*{Conflict of Interest}

The authors declare no conflict of interest

\section*{Author Contributions}
Mi.B., E.Z. conceived the research. E.Z. managed it. Mi.B. carried out the MBE growth, RHEED, AFM, and SEM. M.D. and J.S. performed the \textit{in-situ} Raman measurements. F.R. conducted the EDX experiments, P.A. the XRD experiments.
Mi.B. analyzed the data and discussed the results with all authors. 
H.R., M.D., Mi.B., and E.Z. built the all-UHV MBE-Analytical system and maintained it together with J.S. and A.U. 
J.B. prepared and performed the STEM experiments together with Ma.B. K.V. supervised them.
G.K. and J.J.F. secured third-party funding and provided experimental infrastructure.
Mi.B. wrote the manuscript with input from all authors.

\section*{Data Availability Statement}
The data that support the findings of this study are available from the corresponding author upon reasonable request.

\def\url#1{}

%

\end{document}